\documentclass[onecolumn, draftcls, 12pt]{IEEEtran}

\usepackage{amsfonts}
\usepackage{mathrsfs}
\usepackage{amssymb,amsmath}

\usepackage{algorithm}
\usepackage{algorithmic}
\usepackage{amsmath,amssymb,epsfig,graphics,subfigure}
\usepackage{theorem}
\usepackage{array,color}
\usepackage[compress]{cite}

\theoremheaderfont{\normalfont\bfseries}

\begin{document}

\title{How to Understand LMMSE Transceiver Design for MIMO Systems From Quadratic Matrix Programming}
\author{Chengwen Xing$^*$, Shuo Li, Zesong Fei, and Jingming Kuang
\thanks{Chengwen Xing, Shuo Li, Zesong Fei and Jingming Kuang are with the School of Information and Electronics, Beijing Institute of Technology, Beijing, 100081, China, Phone : (86)
010 6891 1841, (E-mail: chengwenxing@ieee.org, \{surelee,feizesong, jmkuang\}@bit.edu.cn).}
\thanks{The material in this paper was partially presented at the International Conference on Wireless Communications and Signal
Processing (WCSP), Huangshan, China, Sep. 2012. $^*$The corresponding author is
Chengwen Xing.}}

\maketitle

\begin{abstract}
In this paper, a unified linear minimum mean-square-error (LMMSE) transceiver design framework is investigated, which is suitable for a wide range of wireless systems. The unified design is based on an elegant and powerful mathematical programming technology termed as quadratic matrix programming (QMP). Based on QMP it can be observed that for different wireless systems, there are certain common characteristics which can be exploited to design LMMSE transceivers e.g., the quadratic forms. It is also discovered that evolving from a point-to-point MIMO system to various advanced wireless systems such as multi-cell coordinated systems, multi-user MIMO systems, MIMO cognitive radio systems, amplify-and-forward MIMO relaying systems and so on, the quadratic nature is always kept and the LMMSE transceiver designs can always be carried out via iteratively solving a number of QMP problems. A comprehensive framework on how to solve QMP problems is also given. The work presented in this paper is likely to be the first shot for the transceiver design for the future ever-changing wireless systems.
\end{abstract}


\section{Introduction}

In order to satisfy the ever-increasing wireless data rate requirements and to enable high quality and highly diversified wireless services, wireless research never stops to search new discoveries and development in ideas, technologies, systems and everything available. More and more available wireless resources are introduced into wireless systems. The scope of wireless designs has been extended to be multi-dimensional such as temporal, frequency, spatial even coding. As a gift the multi-dimensional wireless resources bring new challenges into wireless system designs. To order to realize the promised performance gains coming from these resources, some corresponding new technologies need to be adopted, such as multiple-carrier technology, multiple-antenna technology and so on.

 Referring to the spatial resource, multiple-input multiple-output (MIMO) technology is a great success in both theoretical research and industrial productions \cite{Telatar1999}. Along with the evolvement of wireless systems, MIMO becomes to be a fundamental and important ingredient of complicated wireless systems e.g., cooperative communications, cognitive communications, physical layer security communications, network coding based communications and so on. Although MIMO technology has promised great potentials in diversity and multiplexing gains, a complicated transmit/receive beamforming or transceiver design is usually needed \cite{Tse2005}. Different from the simple single antenna case, for MIMO transmissions the resources should be carefully allocated across spatial domain according to available channel state information (CSI) at transmitter or receiver or both \cite{Palomar03}.

For MIMO transceiver designs, there are various performance metrics such as capacity, bit error rate (BER), mean-square-error (MSE) and so on. Different performances represent the different preferences of the wireless designers. Meanwhile, because of a variety of wireless service requirements and wireless environments, different wireless systems have totally different network architectures and wireless interfaces. In the resulting transceiver designs, all of these facts are reflected on the constraints and the variables involved in the considered optimization problems. In other words, for different wireless systems the transceiver design problems have different signal models, different power constraints, different numbers of variables and even different performance criteria. As a result transceiver designs must be investigated case by case. From the theoretical research perspective, the theorists and researchers would like to find a unified design which can reveal some common nature of the transceiver designs. To the best our knowledge, up to date the transceiver designs have not been unified for both different performance metrics and different systems. Although the transceiver designs with different performance metrics for different systems are totally different, the work on unifying transceiver designs never stops.

In the existing work for unified linear MIMO transceiver designs, the widely used logic is for a given wireless system linear transceiver designs with different performance metrics are unified into one kind of optimization problem \cite{Palomar03,Sampth01}. It is well-known that there are two guidelines for linear MIMO transceiver designs, i.e., using majorization theory \cite{Palomar03} and weighting operation \cite{Sampth01}. For the majorization theory based guideline, the transceiver design logic is to formulate different performance metrics as different functions of the diagonal elements of the data detection MSE matrix at the destination. Then the objective functions are classified into Schur-convex or Schur-concave functions. Relying on the fundamental properties
of Schur-convex/concave functions, the optimal solutions can be derived. On the other hand, using weighting operations, the different performance metrics are optimized by solving a weighted MSE minimization problem with different weighting matrices.

In this paper, in contrast to the existing work we give a unified transceiver design which aims at unifying the linear transceiver designs for different wireless systems with the same performance metric named as minimum mean-square-error (MMSE). It can be revealed that for the beamforming designs in different wireless systems such as multi-cell coordinated beamforming design, multi-user MIMO beamforming design, cognitive MIMO beamforming design, amplify-and-forward MIMO relaying beamforming design and their corresponding robust transceiver designs with randomly distributed channel estimation errors and so on, the transceiver design problems can always be solved by iteratively solving a series of matrix quadratic programming (QMP) problems that can be efficiently solved.

It is true that our work focuses on iterative linear minimum mean-square-error (LMMSE) transceiver designs which may not be the optimal strategy. This kind of transceiver design suffers from some well-know weaknesses coming from the MMSE objective or iterative design procedure itself or both. We want to highlight that iterative LMMSE designs still have several attractive properties to make them much powerful in engineering applications, as they can be applied to a wide range of fields. Furthermore, they can give a solution with satisfactory performance and they can also act as a benchmark for other kinds of suboptimal schemes.

Wireless systems change very fast e.g., from a point-to-point system to cognitive radio networks or cooperative networks. Although we can describe some about what the future wireless systems is like to be, unfortunately we never know what they are exactly. However, the authors believe that there is definitely something that will not change. Quadratic forms which widely exist are most likely to be kept in transceiver designs because most energy related problems will have quadratic forms. Inspired by this fact, the framework proposed in this paper may work as the first shot which we can do for the coming wireless systems. We also want to highlight that although only transceiver design is investigated in our work, there exist several closely related research topics such as training design in channel estimation procedure \cite{FFGao08} or signal reduction in sensor networks \cite{Schizas07}.
Taking signal reduction \cite{Schizas07} as an example, it is exactly the forwarding matrix design for amplify-and-forward (AF) MIMO relaying systems \cite{Guan08}. In addition it is well-known that training design and transceiver design have the same nature. Then it is not surprising that the solution proposed in this paper can also be applied to such kind of closely related topics. By the way, the main difference between this paper and its conference version \cite{XingWCSP2012} is that the detailed explanations, justifications and discussions are given at various points of the paper. In addition, the important numerical simulations are given in this journal version.

This paper is organized as follows. In Section~\ref{Sect_2}, an interesting understanding of the transceiver designs from optimization theory is presented and it shows the evolvement of transceiver designs is just the same as the procedure to make the optimization problems more complicated. In Section~\ref{Sect_3}, a concrete example of linear transceiver design is first given which shows the motivation for iterative algorithms. Meanwhile, the quadratic nature of LMMSE transceiver design is also revealed. How to exploit the quadratic nature is investigated in Section~\ref{Sect_4} and the framework on QMP is discussed as well. In addition, the applications are specified. After that, an extension on robust designs is considered in Section~\ref{Sect_5}. The numerical results is finally presented in Section~\ref{Sect_6}.

\noindent {\textbf{Notations:}} The following notations are used throughout this paper. Boldface
lowercase letters denote vectors, while boldface uppercase letters
denote matrices. The notations  ${\bf{Z}}^{\rm{T}}$, ${\bf{Z}}^{*}$ and ${\bf{Z}}^{\rm{H}}$ denote the transpose, conjugate and
conjugate transpose of the matrix ${\bf{Z}}$, respectively and ${\rm{Tr}}({\bf{Z}})$ is the
trace of the matrix ${\bf{Z}}$. The symbol ${\bf{I}}_{M}$ denotes an
$M \times M$ identity matrix, while ${\bf{0}}_{M,N}$ denotes an $M
\times N$ all zero matrix. The notation ${\bf{Z}}^{1/2}$ is the
Hermitian square root of the positive semi-definite matrix
${\bf{Z}}$, such that ${\bf{Z}}^{1/2}{\bf{Z}}^{1/2}={\bf{Z}}$ and
${\bf{Z}}^{1/2}$ is also a Hermitian matrix. The symbol ${\mathbb{E}}$ denotes statistical expectation operation. The
operation ${\rm{vec}}({\bf{Z}})$ stacks the columns of the matrix
${\bf{Z}}$ into a single vector. The symbol $\otimes$ represents
Kronecker product.

\section{Motivations}
\label{Sect_2}

At the beginning, we would like to discuss why our attention is concentrated on linear minimum mean-square-error (LMMSE) transceiver designs in this paper. However for ceratin performance metrics such as bit error rate (BER) the performance of linear transceivers may be not as good as that of the nonlinear counterparts, linear transceivers are still preferred by practical wireless systems due to their low complexity. On the other hand,
mean-square-error (MSE) is a widely used performance metric for estimation, detection and optimization algorithm designs in wireless systems. It should be pointed out that MSE acting as performance metric suffers from several inherent drawbacks as it is not the ultimate performance metric e.g., capacity and BER. Roughly speaking, MSE can be seen as an approximation of the ultimate performance metrics, although they have very close relationships and particularly in some special cases, they are even equivalent with each other.
The tractability is the main advantage of MSE. For several ultimate performance metrics, their formulations may be too complicated to optimize. For engineers, the case where there is a solution is much better than that there is no solution.

From the perspective of optimization theory, the LMMSE transceiver designs are in nature some specific optimization problems under different constraints. In general, there are two kinds of variables involved in the optimization problems, i.e., precoder/beamforming matrices and equalizer matrices. The main difference between them is that the equalizers are usually unconstrained. While for precoder/beamforming, the story is different as there are always various kinds of constraints on the transmitters.

The simplest MIMO communication system is the single user point-to-point MIMO system with only one power constraint. The signal model is ${\bf{y}}={\bf{H}}{\bf{F}}{\bf{s}}+{\bf{n}}$ where ${\bf{y}}$ is the received signal at the receiver and ${\bf{H}}$ is the channel matrix between the transmitter and receiver. The symbol ${\bf{F}}$ denotes the precoder matrix at the source, ${\bf{s}}$ is the transmitted signal and ${\bf{n}}$ is the additive noise at the receiver.
The corresponding LMMSE transceiver design problem is formulated as
\begin{align}
\label{ptop}
& \min \ \ \ \ f({\bf{G}},{\bf{F}})={\mathbb{E}}\{\|{\bf{G}}{\bf{y}}-{\bf{s}}\|^2\} \nonumber \\
& \ {\rm{s.t.}} \ \ \ \ \  {\rm{Tr}}({\bf{F}}{\bf{F}}^{\rm{H}}) \le P
\end{align}where ${\bf{G}}$ is the equalizer matrix at the receiver and $P$ represents the maximum transmit power at the transmitter. In the following, we try to understand transceiver designs for various wireless systems evolving from the previous one for the point-to-point MIMO systems.

There are only two possible directions to make the transceiver design problem (\ref{ptop}) more complicated, i.e., enlarging the set of variables or enlarging the sect of constraints. When there are more than one constraint, these constraints can be homogeneous or not (have the same physical meaning or not). As previously discussed, the constraints are always related to transmitters. If the constraints are homogeneous, it means that there may exist many transmitters in the considered wireless system, such as multi-user MIMO (MU-MIMO) uplink. In this case the constraints are described as the second order term of the matrices variables is smaller than a threshold. Of course, the involved constraints can be inhomogeneous. For example, in the cognitive radio, there are usually two kinds of constraints. One is the power constraints and the other is interference constraints. In the latter one, the second order term of the matrices variables is also smaller than a threshold. Different from power constraints it describes that the caused interference in a certain direction must be lower than a threshold. In the following, we refer to the previous kind of constraints with second term smaller than a threshold as positive constraints.

An interesting question is what about the constraint for which the quadratic term is larger than a threshold. It means in a certain direction the energy should be larger than a threshold. In a long time, there is no such kind of wireless systems. Recently, energy harvesting communications give a very important application of this case \cite{Ho2011}. In an energy harvesting communication, except a traditional receiver, there also exists an energy harvesting receiver which aims at harvesting the energy emitted by the transmitter to charge its own battery. As a result, the transmitter should guarantee the energy harvested by the energy harvesting receiver is larger than a threshold. Similar to the case of cognitive radio, in the following we refer to this class of constraints with second term larger than a threshold as negative constraints.

In conclusion, different mathematical formulations of the constraints represent different communication system setups. In the following, we list several concrete and representative examples to illustrate the relationships between various advanced wireless systems and the simplest point-to-point MIMO system. In particular, we want to show how to change of the optimization problem (\ref{ptop}) to become the corresponding optimization problem for an advanced wireless system.

\noindent \textbf{Case 1:} When only the number of  unconstrained variables increases, it corresponds to  MU-MIMO downlink transceiver designs \cite{Zhang05,Serbetli04}. In the following, $f(\bullet)$ represents  the sum MSE function whose specific formulation is determined by the corresponding system model. The linear transceiver design for MU-MIMO downlink is given as follows
\begin{align}
& \min \ \ \ \ f([{\bf{G}}_1,\cdots,{\bf{G}}_K],{\bf{F}}) \nonumber \\
& \ {\rm{s.t.}} \ \ \ \ \  {\rm{Tr}}({\bf{F}}{\bf{F}}^{\rm{H}}) \le P
\end{align}where ${\bf{F}}$ is the beamforming matrix at the base station, and ${\bf{G}}_k$ is the equalizer matrix at the $k^{\rm{th}}$ mobile user. Additionally, $P$ denotes the maximum transmit power at the base station.

\noindent \textbf{Case 2:} When both the numbers of constrained variables and their corresponding constraints increase and the constraints are independent with each other, the result corresponds to MU-MIMO uplink transceiver designs \cite{Larsson03}. In this case, the optimization problem is formulated as
\begin{align}
& \min \ \ \ \ f({\bf{G}},[{\bf{F}}_1,\cdots,{\bf{F}}_K]) \nonumber \\
& \ {\rm{s.t.}} \ \ \ \ \  {\rm{Tr}}({\bf{F}}_k{\bf{F}}_k^{\rm{H}}) \le P_k
\end{align} where ${\bf{F}}_k$ denotes the precoding matrix at the $k^{\rm{th}}$ mobile user and the equalizer at the base station is denoted as ${\bf{G}}$. In addition, $P_k$ denotes the maximum transmit power at the $k^{\rm{th}}$ mobile use.

\noindent \textbf{Case 3:} When both the numbers of constrained variables and unconstrained variables increase and the constraints are independent as well, this case corresponds to multi-cell transceiver designs \cite{XingICASSP2010}. The beamforming design problem for multi-cell cooperation  reads as
\begin{align}
& \min \ \ \ \ f([{\bf{G}}_1,\cdots,{\bf{G}}_K],[{\bf{F}}_1,\cdots,{\bf{F}}_K]) \nonumber \\
& \ {\rm{s.t.}} \ \ \ \ \  {\rm{Tr}}({\bf{F}}_k{\bf{F}}_k^{\rm{H}}) \le P_k,
\end{align}where ${\bf{G}}_k$ is the equalizer at the $k^{\rm{th}}$ base station and ${\bf{F}}_k$ is the precoder matrix at the $k^{\rm{th}}$ mobile terminal.
Moreover, $P_k$ is the maximum transmit power at the $k^{\rm{th}}$ mobile terminal.

\noindent \textbf{Case 4:} Only increase the number of constraints and keep the set of variables unchanged. If the constraints are positive constraint, this case corresponds to cognitive radio (CR) transceiver designs. For CR, the transceiver design problem is formulated as
\begin{align}
& \min \ \ \ \ f({\bf{G}},{\bf{F}}) \nonumber \\
& \ {\rm{s.t.}} \ \ \ \ \  {\rm{Tr}}({\bf{F}}{\bf{F}}^{\rm{H}}) \le P \nonumber \\
& \ \ \ \ \ \ \ \ \ \ {\rm{Tr}}({\bf{H}}_S{\bf{F}}{\bf{F}}^{\rm{H}}{\bf{H}}_S^{\rm{H}}) \le \gamma.
\end{align}where ${\bf{H}}_S$ is the channel matrix between the secondary user node and the primary user node and $\gamma$ denotes the allowable interference threshold.

\noindent \textbf{Case 5:} In contrast to Case 4, when only increasing  negative constraints, it corresponds to energy harvesting oriented transceiver designs. The energy harvesting beamforming design is given as
\begin{align}
& \min \ \ \ \ f({\bf{G}},{\bf{F}}) \nonumber \\
& \ {\rm{s.t.}} \ \ \ \ \  {\rm{Tr}}({\bf{F}}{\bf{F}}^{\rm{H}}) \le P \nonumber \\
& \ \ \ \ \ \ \ \ \ \ {\rm{Tr}}({\bf{H}}_P{\bf{F}}{\bf{F}}^{\rm{H}}{\bf{H}}_P^{\rm{H}}) \ge \gamma,
\end{align}where ${\bf{H}}_P$ denotes the channel between the source node and the energy harvesting node. The physical meaning of the second constraint is in the information transmission, the source node wants to charge the energy harvesting node as well. It should be pointed out that the main difference between Cases 4 and 5 is that the increased constraint is negative or  positive.

\noindent \textbf{Case 6:} When both the number of the constrained variables and the number of corresponding number of constraints increase and meanwhile the constraints are coupled, this case corresponds to the amplify-and-forward (AF) MIMO relaying transceiver designs \cite{XingSP2012}. The transceiver design for two-hop AF MIMO relaying systems can be formulated as
\begin{align}
& \min \ \ \ \ f({\bf{G}},[{\bf{F}}_1,{\bf{F}}_2]) \nonumber \\
& \ {\rm{s.t.}} \ \ \ \ \  {\rm{Tr}}({\bf{F}}_1{\bf{F}}_1^{\rm{H}}) \le P_1 \nonumber \\
& \ \ \ \ \ \ \ \ \ \  {\rm{Tr}}
({\bf{F}}_2({\bf{H}}_1{\bf{F}}_1{\bf{F}}_1^{\rm{H}}
{\bf{H}}_1^{\rm{H}}+\sigma_{n_1}^2{\bf{I}}){\bf{F}}_2^{\rm{H}})\le P_2
\end{align}where ${\bf{F}}_1$ is the source precoder at the source node and ${\bf{F}}_2$ is the forwarding matrix at the relay. Furthermore, ${\bf{H}}_1$ is the channel matrix between the source node and the relay node. In addition $P_1$ and $P_2$ are the maximum transmit power at the source and relay, separately. Notice that the matrix $\sigma_{n_1}^2{\bf{I}}$ is the noise covariance matrix at the relay and ${\bf{H}}_1{\bf{F}}_1{\bf{F}}_1^{\rm{H}}
{\bf{H}}_1^{\rm{H}}+\sigma_{n_1}^2{\bf{I}}$ is the received signal correlation matrix at the relay. It is obvious that the two constraints are coupled with each other. Inspired by the formulation, for a more general multi-hop model, the transceiver design problem becomes
\begin{align}
& \min \ \ \ \ f({\bf{G}},[{\bf{F}}_1,\cdots,{\bf{F}}_K]) \nonumber \\
& \ {\rm{s.t.}} \ \ \ \ \  {\rm{Tr}}({\bf{F}}_1{\bf{F}}_1^{\rm{H}}) \le P_1 \nonumber \\
& \ \ \ \ \ \ \ \ \ \  {\rm{Tr}}({\bf{F}}_k{\bf{R}}_{k-1}{\bf{F}}_k^{\rm{H}})\le P_k \ \ \ \ 2 \le k \le K\nonumber \\
& \ \ \ \ \ \ \ \ \ \  {\bf{R}}_{k-1}={\bf{H}}_{k-1}{\bf{F}}_{k-1}{\bf{R}}_{k-2}
{\bf{F}}_{k-1}^{\rm{H}}{\bf{H}}_{k-1}^{\rm{H}}
+\sigma_{n_{k-1}}^2{\bf{I}} \ \ \ \  2 \le k \le K
\nonumber \\
& \ \ \ \ \ \ \ \ \ \ {\bf{R}}_{0}={\bf{I}}
\end{align}where ${\bf{F}}_k$ is the forwarding matrix at the $k^{\rm{th}}$ node, $P_k$ is the corresponding maximum transmit power and ${\bf{H}}_k$ is the $k^{\rm{th}}$ hop channel matrix. The matrix $\sigma_{n_{k-1}}^2{\bf{I}}$ is the covariance matrix of the additive noise at the $(k-1)^{\rm{th}}$ relay and ${\bf{R}}_{k-1}$ is the received signal correlation matrix at the $(k-1)^{\rm{th}}$ relay .

From Case 1 to Case 6, it can be concluded that the evolvement of wireless communication systems is exactly the evolvement of optimization problem becoming complicated. Of course, the story can continue and we will have Case 7, Case 8 and so on. For engineers, physical meaning is more important than mathematics itself. However, as engineering problems must be perceptible in mathematics here based on these examples we can say that physical meanings cannot be independent of mathematics which can help us to predict what the future communication systems would like to be.

In the following, we will show in detail that for the above optimization problems when iterative algorithms are used, the considered optimization problem admits quadratic nature. As a result, the quadratic matrix programming technology can be used.

\section{Quadratic Nature of the Transceiver Designs}
\label{Sect_3}

In this section, the quadratic nature of the aforementioned optimization problems is investigated. It is totally redundant to discuss it case by case. For simplicity we take \textbf{a representative example} to illustrate that quadratic matrix programming (QMP) problems are of great importance in LMMSE transceiver designs. Note that this example has been discussed in detail in our previous work \cite{MaMILCOM2010}. Here, it only  provides a prologue of our work in this paper. First, we want to  highlight that the algorithm discussed in the following is not limited to this example, which has a much wider application range. We aim at providing a comprehensive framework on LMMSE transceiver design. Our discussions are not limited to any specific communication system. We try to reveal the nature of LMMSE transceiver designs and answer the questions why QMP should always be chosen and how to solve the transceiver design optimization problems using QMP.

\subsection{An example:}
\begin{figure}[!ht]
\centering
\includegraphics[width=.54\textwidth]{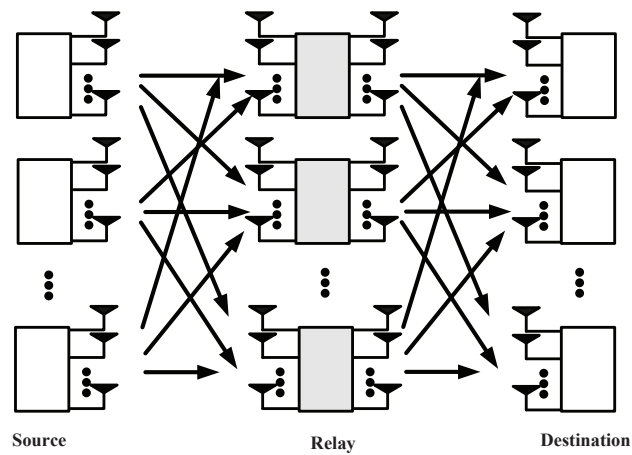}
\caption{A distributed AF MIMO relaying network.}\label{fig:1}
\end{figure}

The considered example is a mixture of Case 3 and Case 6. Here, a dual-hop AF relaying network is investigated. As shown in Fig.~\ref{fig:1}, there are multiple source
nodes, relay nodes and destination nodes. Furthermore, different
sources can have different numbers of transmit antennas and data
streams to transmit. It is denoted that the number of transmit
antennas of the $i^{\rm{th}}$ source is $N_{S,i}$. It is also assumed that for each source node there may be
more than one corresponding destination node. There are also
multiple relay nodes in the network, and the $j^{\rm{th}}$ relay has
$M_{R,j}$ receive antennas and $N_{R,j}$ transmit antennas. At the first hop, the source nodes transmit
data to the relay nodes. The received signal ${\bf{x}}_j$ at the
$j^{\rm{th}}$ relay node is
\begin{align}
{\bf{x}}_{j}&={\bf{H}}_{sr,ij}{\sum}_k({\bf{P}}_{ik}{\bf{s}}_{ik})+{\sum}_{l \ne i}[{\bf{H}}_{sr,lj}{\sum}_k({\bf{P}}_{lk}{\bf{s}}_{lk})]\nonumber \\
&+{\bf{n}}_{1,j}.
\end{align}
where ${\bf{s}}_{ik}$ is the data vector transmitted by the
$i^{\rm{th}}$ source node to the $k^{\rm{th}}$ destination with the
covariance matrix
${\bf{R}}_{{\bf{s}}_{ik}}=\mathbb{E}\{{\bf{s}}_{ik}{\bf{s}}_{ik}^{\rm{H}}\}$.
When the $i^{\rm{th}}$ source node does not want to transmit signal
to the $k^{\rm{th}}$ destination, ${\bf{s}}_{ik}$ is a all-zero vector.

At the source, before transmission the signal is multiplied a
precoder ${\bf{P}}_{ik}$ under al transmit power constraint
$\sum_k{\rm{Tr}}({\bf{P}}_{ik}{\bf{R}}_{{\bf{s}}_{ik}}{\bf{P}}_{ik}^{\rm{H}})\le
P_{s,i}$, where $P_{s,i}$ is the maximum transmit power at the
$i^{\rm{th}}$ source node. The matrix ${\bf{H}}_{sr,ij}$ is the MIMO
channel matrix between the $i^{\rm{th}}$ source node and the
$j^{\rm{th}}$ relay node. Symbol ${\bf{n}}_{1,j}$ is the additive
Gaussian noise with the covariance matrix
${\bf{R}}_{{\bf{n}}_{1,j}}$. At the $j^{\rm{th}}$ relay node, the
received signal ${\bf{x}}_{j}$ is multiplied by a precoder matrix
${\bf{F}}_j$, under a power constraint
${\rm{Tr}}({\bf{F}}_j{\bf{R}}_{{\bf{x}}_j}{\bf{F}}_j^{\rm{H}}) \le
P_{r,j}$ where
${\bf{R}}_{{\bf{x}}_j}=\mathbb{E}\{{\bf{x}}_j{\bf{x}}_j^{\rm{H}}\}$
and $P_{r,j}$ is the maximum transmit power. Then the resulting
signal is transmitted to the destination. The received signal at the
$k^{\rm{th}}$ destination ${\bf{y}}_k$ can be written as
\begin{align}
\label{signal_destin}
{\bf{y}}_{k} & ={\sum}_{j}({\bf{H}}_{rd,jk}{\bf{F}}_{j}{\bf{x}}_{j})+{\bf{n}}_{2,i} \nonumber \\
&={\sum}_{j}[{\bf{H}}_{rd,jk}{\bf{F}}_j{\sum}_l({\bf{H}}_{sr,lj}{\bf{P}}_{lk}{\bf{s}}_{lk})]
\nonumber \\
&+{\sum}_j[{\bf{H}}_{rd,ji}{\bf{F}}_j{\sum}_l({\bf{H}}_{sr,lj}{\sum}_{m \ne k }({\bf{P}}_{lm}{\bf{s}}_{lm}))]\nonumber \\
&+{\sum}_j({\bf{H}}_{rd,jk}{\bf{F}}_j{\bf{n}}_{1,j}) +{\bf{n}}_{2,k}.
\end{align} where ${\bf{H}}_{rd,jk}$ is the MIMO channel matrix between the $j^{\rm{th}}$ relay
and the $k^{\rm{th}}$ destination, and ${\bf{n}}_{2,k}$ is the additive Gaussian noise
vector at the $k^{\rm{th}}$ destination with covariance matrix ${\bf{R}}_{{\bf{n}}_{2,k}}$.

The optimization problem of linear minimum mean-square-error (LMMSE) transceiver design can be formulated as \cite{MaMILCOM2010}
\begin{align}
\label{prob:opt}
& \min \ \ \ {\sum}_k {\rm{MSE}}_k={\mathbb{E}}\{\|{\bf{G}}_k{\bf{y}}_k-[{\bf{s}}_{1k}^{\rm{T}},\cdots,{\bf{s}}_{N_sk}^{\rm{T}}]^{\rm{T}}\|^2\} \nonumber \\
& \ {\rm{s.t.}} \ \ \ \ {\rm{Tr}}({\bf{F}}_j{\bf{R}}_{{\bf{x}}_j}{\bf{F}}_j^{\rm{H}})\le P_{r,j} \ \ j\in {\mathcal{E}}_{r} \nonumber \\
& \ \ \ \ \ \ \ \ \ {\sum}_k {\rm{Tr}}({\bf{P}}_{ik}{\bf{R}}_{{\bf{s}}_{ik}}{\bf{P}}_{ik}^{\rm{H}} )\le P_{s,i} \ \ i\in {\mathcal{E}}_{s}
\end{align}where $[{\bf{s}}_{1k}^{\rm{T}},\cdots,{\bf{s}}_{N_sk}^{\rm{T}}]^{\rm{T}}$ is the desired signal to be recovered at the $k^{\rm{th}}$ destination. Additionally ${\mathcal{E}}_{r} $ and ${\mathcal{E}}_{s}$ denote the set of relay nodes and the set of source nodes, respectively.

The optimization problem (\ref{prob:opt}) is a very general problem which includes the following scenarios as its special cases.

\noindent $\bullet$  Multi-user MIMO uplink transceiver design  \cite{XingSP2012}: Multiple multi-antenna mobile users communicate with a multi-antenna base station.

\noindent $\bullet$  Multi-user MIMO downlink transceiver design \cite{XingSP2012}: A multi-antenna base station communicates with multiple multi-antenna mobile users.

\noindent $\bullet$ Multi-cell coordinated beamforming design: Multiple multi-antenna base stations communicate cooperatively with multiple multi-antenna mobile users.

\noindent $\bullet$ Two-way AF MIMO relaying LMMSE transceiver design \cite{XingICSPCC2011}: Two-way AF MIMO relaying can be taken as a soft combination of uplink and downlink beamforming designs. Although, the optimization problem (\ref{prob:opt}) only considers one-way relaying systems. The extension from one-way to two-way is straightforward when an iterative optimization framework is used.

\subsection{Iterative Algorithms}

As in the optimization (\ref{prob:opt}) there are too many variables to be optimized and meanwhile the nonconvex nature of the optimization problem (\ref{prob:opt}) makes it very complicated, generally it is difficult to find the closed-form globally optimal solutions. In order to design the transceivers, several suboptimal solutions are usually proposed. Iterative algorithm is one of the most widely used and important suboptimal solutions. In an iterative algorithm, the variables are optimized sequentially.
It can be interpreted that iterative algorithms use iterative procedure to soften the hardness of the original optimization problems as in the iterative procedure the coupling relationships among the involved variables can be removed first.

We admit that iterative algorithms suffer from some well-known weaknesses. First, the final solution is greatly affected by the initial value selection. Second, the convergence of an iterative algorithm must be guaranteed. If not, the algorithm may be meaningless.
Third, in general even with proved convergence there is no guarantee that the final solution is globally optimal. However, iterative algorithms still have two important characteristics making them preferable. First, it can be applied to a much wide area of transceiver designs ranging from a point-to-point system to a distributed network. Second, it can act as a performance benchmark for other suboptimal solutions. Actually iterative algorithms are widely adopted in transceiver designs or beamforming designs for MIMO systems no matter you love it or hate it \cite{Choi2010}. When iterative algorithms are adopted to solve the optimization problem (\ref{prob:opt}), in each iteration one variable is optimized and the others are fixed, and then the problem admits quadratic nature.

\subsection{Quadratic nature of the LMMSE transceiver designs}

Data detection MSE is an integration over the signals and noises. From its name, it is obvious that MSE is a certain quadratic formulation with respect to each involved variable. Moreover, in this paper, we concentrate our attention to the case where the variables are matrices, as in MIMO systems the variables to be optimized are usually complex matrices. Inspired by these facts, to characterize a quadratic function with a complex matrix variable a kind of functions termed as quadratic matrix (QM) functions  ${\bf{X}}$ is first defined as
\begin{align}
f_l({\bf{X}})={\rm{Tr}}({\bf{D}}_l{\bf{X}}^{\rm{H}}{\bf{A}}_l{\bf{X}})+2{\mathcal R}\{{\rm{Tr}}({\bf{B}}_l^{\rm{H}}{\bf{X}})\}+c_l
\end{align}where ${\bf{A}}_l={\bf{A}}_l^{\rm{H}} \in {\mathbb C}^{n\times n}$, ${\bf{B}}_l\in {\mathbb C}^{n\times r}$, $c_l \in {\mathbb R}$, ${\bf{D}}_l={\bf{D}}_l^{\rm{H}} \in {\mathbb C}^{r\times r}$. In addition, ${\mathcal R}\{\bullet\}$ denotes the real part. It can be seen that a QM function consists of three terms which are second-order term, first-order term and zero-order term. If the following conditions are satisfied, not matter what the system is, the MSE with linear transceivers is a QM function with respect to each variable, separately.

\noindent (1). The considered system is a linear system. Linearity is defined based on the following two properties:

(a.1) The received signal at the destination is a linear function of the transmit signal when all design variables are fixed.

(a.2) The received signal at the destination is a linear function with respect to each variable when the signal and the other design variables are all fixed.

\noindent (2). The desired signals are independent of the noises.
 It means that when the transmit signal vector is denoted by ${\bf{s}}$ and the equivalent noise vector is ${\bf{v}}$, the following equality must hold
 \begin{align}
 \mathbb{E}\{{\bf{s}}{\bf{v}}^{\rm{H}}\}={\bf{0}}.
 \end{align}

Moreover, the constraints in the transceiver designs for wireless systems are usually QM functions as well. This is because the involved constraints are usually related with energy, which definitely have quadratic terms e.g., transmit power, interference to primary users, and so on. Therefore, it is of great importance to investigate the optimization problems consisting of QM functions in both objective function and constraint functions. This kind of optimization problems is named as quadratic matrix programming (QMP) problems. It can be observed that in each iteration, the optimization problem (\ref{prob:opt}) becomes a QMP problem. Although in \cite{Beck07}, a definition of quadratic matrix programming is given, in this paper we first revise the definition given in \cite{Beck07} in order to accommodate more cases e.g., the problem (\ref{prob:opt}). As a result, our definition is more general and has a wider range of applications. A standard QMP problem is defined as
\begin{align}
\label{optimization_P_1}
& \textbf{Type 1 QMP:} \nonumber \\
& \min_{\bf{X}} \ \ {\rm{Tr}}({\bf{D}}_0{\bf{X}}^{\rm{H}}{\bf{A}}_0{\bf{X}})+2{\mathcal R}\{{\rm{Tr}}({\bf{B}}_0^{\rm{H}}{\bf{X}})\}+c_0 \nonumber \\
& \ {\rm{s.t.}} \ \ \ {\rm{Tr}}({\bf{D}}_i{\bf{X}}^{\rm{H}}{\bf{A}}_i{\bf{X}})+2{\mathcal R}\{{\rm{Tr}}({\bf{B}}_i^{\rm{H}}{\bf{X}})\}+c_i \le 0, i \in {\mathcal I} \nonumber \\
& \ \ \ \ \ \ \ \ {\rm{Tr}}({\bf{D}}_j{\bf{X}}^{\rm{H}}{\bf{A}}_j{\bf{X}})+2{\mathcal R}\{{\rm{Tr}}({\bf{B}}_j^{\rm{H}}{\bf{X}})\}+c_j=0, j \in {\mathcal E} \nonumber \\
& \ \ \ \ \ \ \ \ {\bf{X}} \in {\mathbb C}^{n\times r}
\end{align}where ${\bf{A}}_l={\bf{A}}_l^{\rm{H}} \in {\mathbb C}^{n\times n}$, ${\bf{B}}_l\in {\mathbb C}^{n\times r}$, $c_l \in {\mathbb R}$, ${\bf{D}}_l={\bf{D}}_l^{\rm{H}} \in {\mathbb C}^{r\times r}$, $l \in \{0\} \cup {\mathcal{I}} \cup {\mathcal {E}}$. These assumptions are essential to guarantee that the objective function and constraint functions are real-valued functions, as it is meaningless to minimize a complex-valued function. The main difference between our definition and that given in \cite{Beck07} is that in \cite{Beck07}  ${\bf{D}}_l={\bf{I}}$ while in our definition they can be arbitrary Hermitian matrices. In the following section, the important characteristics of QMP problems will be discussed, based on which a comprehensive framework on how to solve it is also given. In the sequel the Type 1 QMP problems
are abbreviated to be T-1-QMP problems.

\section{Fundamentals of QMP}
\label{Sect_4}

In this section, the fundamental properties of QMP are investigated. It is obvious that quadratic matrix programming (QMP) is a special case of quadratically constrained quadratic programming (QCQP) which is a very famous and widely used  \cite{ZQLuo2010}. Obviously the QMP problems have much better properties (e.g., Kronecker structure) than traditional QCQP problems, which can be further exploited to solve the considered optimization problems more efficiently. This is exactly the motivation of the research on QMP \cite{Beck07,Beck09}.
We embark on our investigation from the T-1-QMP
problems in (\ref{optimization_P_1}), which are the most  general problems.

\noindent \textbf{General QMP:}

Based on the
properties of Kronecker product and the following definitions
\begin{align}
{\boldsymbol \Omega}_l \triangleq \left[ {\begin{array}{*{20}c}
  {\bf{D}}_{l}^{\rm{T}} \otimes {\bf{A}}_l & {\rm{vec}}({{\bf{B}}_l)}  \\
   {\rm{vec}}^{\rm{H}}({\bf{B}}_l) & {c_l}   \\
\end{array}} \right], \ \ l \in \{0\} \cup {\mathcal{I}} \cup {\mathcal {E}}
\end{align}the optimization problem (\ref{optimization_P_1}) is equivalent to
\begin{align}
\label{SDP}
& {\min} \ \ \ {\rm{Tr}}({\boldsymbol \Omega}_0{\bf{Z}}) \nonumber \\
& {\rm{s.t.}} \ \ \ \ {\rm{Tr}}({\boldsymbol \Omega}_i{\bf{Z}}) \le
0, \ \ {\rm{Tr}}({\boldsymbol \Omega}_j{\bf{Z}})=0
\nonumber \\
&\ \ \ \ \ \ \ \  {\bf{Z}}=[{\rm{vec}}^{\rm{T}}({\bf{X}}) \
1]^{\rm{T}}[{\rm{vec}}^{\rm{H}}({\bf{X}}) \ 1].
\end{align}If the constraint ${\rm{Rank}}({\bf{Z}})=1$ is relaxed
(it is a well-known semi-definite relaxation (SDR) \cite{Beck09}), we have the following semi-definite programming (SDP) problem \cite{Vandenberghe96},  which can be efficiently solved by
interior point polynomial algorithms
\begin{align}
\label{SDR}
& {\min\limits_{\bf{Z}}} \ \ \ {\rm{Tr}}({\boldsymbol \Omega}_0{\bf{Z}}) \nonumber \\
& {\rm{s.t.}} \ \ \ \ {\rm{Tr}}({\boldsymbol \Omega}_i{\bf{Z}}) \le
0, \ \  {\rm{Tr}}({\boldsymbol \Omega}_j{\bf{Z}})=0
\nonumber \\
&\ \ \ \ \ \ \ \  [{\bf{Z}}]_{NN_s+1,NN_s+1}=1, \ \ {\bf{Z}}
\succeq 0,
\end{align}where ${\bf{Z}}$ is a Hermitian matrix.

\noindent \textbf{Applications:} Generally speaking, for iterative LMMSE transceiver designs for the previously considered systems  in each iteration the variables can always be solved using the solution for the general QMP problem, e.g, multi-cell transceiver designs, CR transceiver designs, energy harvesting transceiver designs, AF MIMO relaying transceiver designs and so on.

\noindent \textbf{Convex QMP:}
When ${\bf{A}}_l$ and ${\bf{D}}_l$ are both positive semi-definite matrices and the involved constraints are only inequality constraints, the QMP problem (\ref{optimization_P_1}) is convex \cite{Boyd04}.
Convexity may be the most favorable property for an optimization problem and convex optimization problems can usually be efficiently solved. In the sequel, it is revealed that for convex QMP problems, it does not need the previous SDR to compute the optimal solutions. In the following, two approaches to solving convex QMP problems are proposed.

\noindent \underline{\textbf{SDP Based Algorithm:}}

Using the properties of Kronecker product ${\rm{Tr}}({\boldsymbol{A}}{\boldsymbol{B}})={\rm{vec}}^{\rm{H}}({\boldsymbol{A}}^{\rm{H}}){\rm{vec}}({\boldsymbol{B}})$, the QM function can be reformulated as
\begin{align} &{\rm{Tr}}({\bf{D}}_l^{\frac{\rm{H}}{2}}{\bf{X}}^{\rm{H}}{\bf{A}}_l
{\bf{X}}{\bf{D}}_l^{\frac{1}{2}})
+2\mathcal{R}\{{\rm{Tr}}({\bf{B}}_l^{\rm{H}}{\bf{X}})\}+c_l \nonumber \\
&={\rm{Tr}}({\bf{D}}_l^{\frac{\rm{H}}{2}}{\bf{X}}^{\rm{H}}{\bf{A}}_l^
{\frac{\rm{H}}{2}}
{\bf{A}}_l^{\frac{1}{2}}
{\bf{X}}{\bf{D}}_l^{\frac{1}{2}})
+2\mathcal{R}\{{\rm{Tr}}({\bf{B}}_l^{\rm{H}}{\bf{X}})\}+c_l \nonumber \\
&={\rm{vec}}^{\rm{H}}({\bf{X}})({\bf{D}}_l^{\frac{*}{2}}\otimes {\bf{A}}_l^{\frac{\rm{H}}{2}})({\bf{D}}_l^{\frac{\rm{T}}{2}}\otimes {\bf{A}}_l^{\frac{1}{2}}){\rm{vec}}({\bf{X}})\nonumber \\
&+2\mathcal{R}\{{\rm{vec}}^{\rm{H}}({\bf{B}}_l){\rm{vec}}({\bf{X}})\}+c_l
\le 0,
\end{align} based on which and together with Schur complement lemma, the optimization problem (\ref{optimization_P_1}) can be reformulated as
\begin{align}
\label{SDR}
&{\min} \ \ \ \ \ \ \ \  t \nonumber \\
& \ {\rm{s.t.}}  \ \
 \left[ {\begin{array}{*{20}c}
   {\bf{I}} & ({\bf{D}}_0^{\frac{\rm{T}}{2}}\otimes {\bf{A}}_0^{\frac{1}{2}}){\rm{vec}}({\bf{X}})   \\
   (({\bf{D}}_0^{\frac{\rm{T}}{2}}\otimes {\bf{A}}_0^{\frac{1}{2}}){\rm{vec}}({\bf{X}}))^{\rm{H}} & -2\mathcal{R}({\rm{vec}}^{\rm{H}}({\bf{B}}_0){\rm{vec}}({\bf{X}}))
   +t  \\
\end{array}} \right]  \succeq 0 \nonumber  \\
& \ \ \ \ \ \ \ \left[ {\begin{array}{*{20}c}
   {\bf{I}} & ({\bf{D}}_i^{\frac{\rm{T}}{2}}\otimes {\bf{A}}_i^{\frac{1}{2}}){\rm{vec}}({\bf{X}})   \\
   (({\bf{D}}_i^{\frac{\rm{T}}{2}} \otimes {\bf{A}}_i^{\frac{1}{2}}){\rm{vec}}({\bf{X}}))^{\rm{H}} & -2\mathcal{R}({\rm{vec}}^{\rm{H}}({\bf{B}}_i){\rm{vec}}
   ({\bf{X}}))-c_i \\
\end{array}} \right] \succeq 0.
\end{align}

Notice that in our work, the variables are complex matrices. For some optimization toolboxes, maybe only real variables are permitted. In that case, only a minor transformation is needed, which is
\begin{align}
\left[ {\begin{array}{*{20}c}
  {\bf{I}}_N &  {\bf{v}} \\
    {\bf{v}}^{\rm{H}}& a  \\
\end{array}} \right] \succeq 0 \rightarrow \left[ {\begin{array}{*{20}c}
  {\bf{I}}_{2N} &  {\bf{\tilde v}} \\
    {\bf{\tilde v}}^{\rm{T}}& a  \\
\end{array}} \right] \succeq 0
\end{align} where $ {\bf{\tilde v}}$ is defined as
\begin{align}
 {\bf{\tilde v}}=[{\rm{Real}}({\bf{ v}})^{\rm{T}} \ {\rm{Imag}}({\bf{ v}})^{\rm{T}}]^{\rm{T}}.
\end{align}

Furthermore, if ${\bf{A}}_i$ and ${\bf{D}}_i$ are both positive definite matrices (stronger than positive semidefinite matrices), the optimization problem can be further transformed into a more efficiently solvable convex optimization problem e.g., second order conic programming (SOCP) problems.

\noindent \underline{\textbf{SOCP Based Algorithm:}}

Notice that when ${\bf{A}}_i$ and ${\bf{D}}_i$ are both positive definite, the QM functions in both the objective function and constraints can be reformulated as
\begin{align}
&{\rm{Tr}}({\bf{D}}_l^{\rm{H}/2}{\bf{X}}^{\rm{H}}{\bf{A}}_l
{\bf{X}}{\bf{D}}_l^{1/2})
+2\mathcal{R}\{{\rm{Tr}}({\bf{B}}_l^{\rm{H}}{\bf{X}})\}+c_l\nonumber \\
=&\left\| \left[{\begin{array}{*{20}c}
   {{\bf{A}}_l^{\frac{1}{2}}{\bf{X}}{\bf{D}}_l^{\frac{1}{2}}
   +{\bf{A}}_l^{-\frac{1}{2}}{\bf{B}}_l{\bf{D}}_l^{-\frac{1}{2}}}  \\
\end{array}}\right] \right\|_{\rm{F}}^2+c_i\nonumber \\
&-{\rm{Tr}}({\bf{A}}_l^{-1}{\bf{B}}_l{\bf{D}}_l^{-1}{\bf{B}}_l^{\rm{H}})
\end{align}where $\|\bullet\|_{\rm{F}}$ denotes  Frobenious norm. Therefore,
the optimization problem (\ref{optimization_P_1}) can be reformulated as a standard SOCP problem which reads as
\begin{align}
& \min_{{\bf{P}}_k,t} \ \ \ t \nonumber \\
& {\rm{s.t.}} \ \ \ \left\| \left[{\begin{array}{*{20}c}
   {{\bf{A}}_0^{\frac{1}{2}}{\bf{X}}{\bf{D}}_0^{\frac{1}{2}}
   +{\bf{A}}_0^{-\frac{1}{2}}{\bf{B}}_0{\bf{D}}_0^{-\frac{1}{2}}}  \\
\end{array}}\right] \right\|_{\rm{F}}\le t \nonumber \\
& \ \ \ \ \ \ \ \left\| \left[{\begin{array}{*{20}c}
   {{\bf{A}}_i^{\frac{1}{2}}{\bf{X}}{\bf{D}}_i^{\frac{1}{2}}
   +{\bf{A}}_i^{-\frac{1}{2}}{\bf{B}}_i{\bf{D}}_i^{-\frac{1}{2}}}  \\
\end{array}}\right]  \right\|_{\rm{F}} \le \sqrt{{\rm{Tr}}({\bf{A}}_i^{-1}{\bf{B}}_i{\bf{D}}_i^{-1}
{\bf{B}}_i^{\rm{H}})-c_i}.
\end{align}

\noindent \textbf{Applications:} Convex-QMP is suitable for multi-cell transceiver designs and AF MIMO relaying transceiver designs.

Based on the previous discussions it can be concluded that the better structure the stronger solution. In the remaining part of this section, we will take a further step to concentrate our attention to the QMP problems which have the following special structure
\begin{align}
\label{optimization_P_2}
& \textbf{Type 2 QMP:} \nonumber \\
& \min_{\bf{X}} \ \ {\rm{Tr}}({\bf{X}}^{\rm{H}}{\bf{A}}_0{\bf{X}})+2{\mathcal R}\{{\rm{Tr}}({\bf{B}}_0^{\rm{H}}{\bf{X}})\}+c_0 \nonumber \\
& \ {\rm{s.t.}} \ \ \ {\rm{Tr}}({\bf{X}}^{\rm{H}}{\bf{A}}_i{\bf{X}})+2{\mathcal R}\{{\rm{Tr}}({\bf{B}}_i^{\rm{H}}{\bf{X}})\}+c_i \le 0, i \in {\mathcal I} \nonumber \\
& \ \ \ \ \ \ \ \ {\rm{Tr}}({\bf{X}}^{\rm{H}}{\bf{A}}_j{\bf{X}})+2{\mathcal R}\{{\rm{Tr}}({\bf{B}}_j^{\rm{H}}{\bf{X}})\}+c_j=0, j \in {\mathcal E} \nonumber \\
& \ \ \ \ \ \ \ \ {\bf{X}} \in {\mathbb C}^{n\times r}.
\end{align} The Type 2 QMP problems are also usually encountered in the LMMSE transceiver designs for wireless communications \cite{Xing1012}. It is worth investigating its properties detailedly. For the notational simplicity, in the following the T-2-QMP problems are referred to as the Type 2 QMP problems.

\subsection{Properties of T-2-QMP}
\subsubsection{T-2-QMP without Constraints}

At the first glance, we discuss the case without constraint which reads as
\begin{align}
& \min_{\bf{X}} \ \ {\rm{Tr}}({\bf{X}}^{\rm{H}}{\bf{A}}_0{\bf{X}})+2{\mathcal R}\{{\rm{Tr}}({\bf{B}}_0^{\rm{H}}{\bf{X}})\}+c_0
\end{align}where ${\bf{A}}_0>{\bf{0}}$. This case corresponds to linear minimum mean square error (LMMSE) equalizer design, which is also named as LMMSE estimator design. In this case, as previously discussed, the optimization problem is convex and the optimal solution is exactly the solution making the differentiation of the objective equation equal 0 i.e., $
{\bf{A}}_0{\bf{X}}=-{\bf{B}}_0 $. Specifically, the optimal solution has the following closed-form solution
\begin{align}
{\bf{X}}_{\rm{opt}}=-{\bf{A}}_0^{-1}{\bf{B}}_0.
\end{align} This solution is a very strong solution, which is also the optimal solution of weighted MSE minimization problem independent of weighting matrices.

Weighted MSE is a direct generalization of sum MSE.
Considering weighted MSE minimization, the optimization problem becomes to be
\begin{align}
& \min_{\bf{X}} \ \ {\rm{Tr}}({\bf{W}}_{\rm{w}}{\bf{X}}^{\rm{H}}{\bf{A}}_0{\bf{X}})+2{\mathcal R}\{{\rm{Tr}}({\bf{W}}_{\rm{w}}^{\rm{H}}{\bf{B}}_0^{\rm{H}}{\bf{X}})\}+c_0
\end{align}where ${\bf{W}}_{\rm{w}}\succeq {\bf{0}}$ is the weighting matrix. Following the same logic as previously discussed, the optimal solutions must satisfy
\begin{align}
\label{weighted MSE optimal solution}
{\bf{A}}_0{\bf{X}}{\bf{W}}_{\rm{w}}=-{\bf{B}}_0{\bf{W}}_{\rm{w}}.
\end{align} Actually, this condition is a sufficient condition for the optimal solution as the optimization problem is convex. Because ${\bf{W}}_{\rm{w}}$ can be ill-rank, the optimal solution is not unique. Notice that the following solution satisfying the previous condition (\ref{weighted MSE optimal solution})
\begin{align}
{\bf{X}}_{\rm{opt}}=-{\bf{A}}_0^{-1}{\bf{B}}_0.
\end{align}This conclusion is important as it shows that ${\bf{X}}_{\rm{opt}}$ is a dominating estimator. It is why even for capacity achieving transceiver designs, LMMSE equalizer is optimal.

\noindent \underline{\textbf{Conclusion 1:}} Without constraints, the optimal solution ${\bf{X}}_{\rm{opt}}$ of the T-2-QMP problems  has a closed form. Notice that ${\bf{X}}_{\rm{opt}}^{\rm{H}}$ is just the Wiener filter. It is well-known for a linear system with Gaussian noise, LMMSE equalizer is exactly the optimal equalizer in the sense of both linear equalizers and nonlinear equalizers \cite{Kay93}. To the best of our knowledge, this solution can be applied to all linear equalizer designs in wireless systems.

\subsubsection{T-2-QMP with One Constraint}

After discussing the case without constraints, we take a step further to focus on the case where there is only one constraint for the considered QMP problem. This case corresponds to the scenario when there is only one transmit power constraint. Here  we focus on the following T-2-QMP problem
\begin{align}
\label{Prob_one}
& \min_{\bf{X}} \ \ {\rm{Tr}}({\bf{X}}^{\rm{H}}{\bf{A}}_0{\bf{X}})+2{\mathcal R}\{{\rm{Tr}}({\bf{B}}_0^{\rm{H}}{\bf{X}})\}+c_0 \nonumber \\
& \ {\rm{s.t.}} \ \ \ {\rm{Tr}}({\bf{X}}^{\rm{H}}{\bf{A}}_1{\bf{X}}) \le P,
\end{align}where ${\bf{A}}_l>{\bf{0}}$. For the problem we considered, the feasible set is not empty. In this scenario,  solving the matrix variable can be reduced to solve an unknown scalar variable. The computational dimensionality and complexity are both significantly reduced. In this following, we will discuss this in detail.

For constrained optimization problems, if certain regularity conditions are satisfied, Karush-Kuhn-Tucker (KKT) are the necessary conditions for the optimal solutions and then KKT conditions can provide very important information to help us find the optimal solutions.
When there is one constraint, linear independence of constraint qualification (LICQ) can be easily proved, which is a famous regularity condition \cite{Bertsekas99,Bertsekas03}. In this case, the condition for LICQ to hold is that the optimal solution ${\bf{X}}$ is not all zero matrix. In practical wireless systems, this is always true as when transmitter matrix is all zero, there is no information to be transmitted and of course it is not the optimal solution. Therefore,  KKT conditions are the necessary conditions for the optimal solutions \cite{Bertsekas99,Bertsekas03}.

The corresponding Lagrange function of the optimization problem (\ref{Prob_one}) is expressed as
\begin{align}
\label{Lagrange_function}
\mathcal{L}({\bf{X}})&={\rm{Tr}}({\bf{X}}^{\rm{H}}{\bf{A}}_0{\bf{X}})+2{\mathcal R}\{{\rm{Tr}}({\bf{B}}_0^{\rm{H}}{\bf{X}})\}+c_0\nonumber \\
& +\mu( {\rm{Tr}}({\bf{X}}^{\rm{H}}{\bf{A}}_1{\bf{X}})-P),
\end{align}where $\mu\ge0$ is the Lagrange multiplier. Based on (\ref{Lagrange_function}),  the KKT conditions of the optimization problem (\ref{Prob_one}) can be directly derived to be \cite{Boyd04}
\begin{align}
&({\bf{A}}_0+\mu{\bf{A}}_1){\bf{X}}=-{\bf{B}}_0 \\
&\mu( {\rm{Tr}}({\bf{X}}^{\rm{H}}{\bf{A}}_1{\bf{X}})-P)=0  \\
&{\rm{Tr}}({\bf{X}}^{\rm{H}}{\bf{A}}_1{\bf{X}}) \le P \\
& \mu\ge0.
\end{align}

In this case with a single constraint, the optimal solution has the following semi-closed-form solution
\begin{align}
\label{X_solution}
{\bf{X}}=-({\bf{A}}_0+\mu{\bf{A}}_1)^{-1}{\bf{B}}_0
\end{align}in which the only unknown variable is a scalar Lagrange multiplier. Substituting (\ref{X_solution}) into the constraint of (\ref{Prob_one}), we have
\begin{align}
&{\rm{Tr}}({\bf{X}}^{\rm{H}}{\bf{A}}_1{\bf{X}})\nonumber \\
=&{\rm{Tr}}
({\bf{B}}_0^{\rm{H}}({\bf{A}}_0+\mu{\bf{A}}_1)^{-1}
{\bf{A}}_1({\bf{A}}_0+\mu{\bf{A}}_1)^{-1}{\bf{B}}_0)\nonumber \\
=&{\rm{Tr}}
({\bf{B}}_0^{\rm{H}}{\bf{A}}_1^{-\frac{1}{2}}({\bf{A}}_1^{-\frac{1}{2}}
{\bf{A}}_0{\bf{A}}_1^{-\frac{1}{2}}+\mu{\bf{I}})
^{-2}{\bf{A}}_1^{-\frac{1}{2}}{\bf{B}}_0) \nonumber \\
\triangleq & g(\mu).
\end{align}It has been proved that $g(\mu)$ is a decreasing function with respect to $\mu$ \cite{XingICASSP2010}, and the value of $\mu$ satisfying the KKT conditions can be found by using a simple one-dimensional search such as bisection search. Based on this conclusion and the KKT conditions given previously, the value of $\mu$ can be computed to be
\begin{equation}
\label{gamma}
\mu=\begin{cases} 0 & \text{if $g(0) \le P$} \\
\text{Solve $g(\mu)=P$ } &
\text{Otherwise}
\end{cases}
.
\end{equation}

It can be seen that the solution satisfying KKT conditions is unique.
As the KKT conditions are the necessary conditions for the optimal solutions. As a result, the unique solution satisfying the KKT conditions is exactly the optimal solution.
This is of great importance. In this case, the unknown variable is simplified from a matrix to a scalar. In other words, the number of variables is significantly reduced and the corresponding computational complexity will be significantly reduced.



\noindent \underline{\textbf{Conclusion 2:}} With only one constraint, the T-2-QMP problem has a semi-closed-form solution with an unknown scalar variable.  This solution is applicable to downlink MU-MIMO beamforming design at the base station and amplifying matrix design for the dual-hop AF MIMO relaying transceiver designs (including both one-way and two-way).

\noindent {\textbf{Remark:}} We cannot argue that KKT conditions are necessary conditions for the optimal solutions without any prior conditions. In Boyd's classical textbook \cite{Boyd04}, it never states that KKT conditions are necessary optimality conditions for any optimization problems. There are several cases in which KKT conditions are not necessary optimality conditions \cite{Bertsekas03}.

\subsubsection{T-2-QMP with more than one constraint}

For T-2-QMP problems with more than one constraint, solving the optimization problems must also rely on interior point algorithms. As a T-2-QMP problem has much better structures comparing to a general QMP problem discussed in the previous section, it exhibits more stronger convexity property which can be exploited to solve the optimization problem.
As discussed in \cite{Beck09}, the original optimization problem is first transformed into its homogenized problem which can be efficiently solved. First, the homogenized QM function of the QM function defined previously is denoted by $f_i^{\rm{H}}$
\begin{align}
f_l^{\rm{H}}({\bf{Y}};{\bf{Z}})=&{\rm{Tr}}({\bf{Y}}^{\rm{H}}{\bf{A}}_l{\bf{Y}})+2{\mathcal R}\{{\rm{Tr}}({\bf{Z}}^{\rm{H}}{\bf{B}}_l^{\rm{H}}{\bf{Y}})\}\nonumber \\
&+\frac{c_l}{r}{\rm{Tr}}({\bf{Z}}^{\rm{H}}{\bf{Z}}).
\end{align}Then introducing the following operators,
\begin{align}
{\bf{M}}_l(f_l)=
\left[ {\begin{array}{*{20}c}
   {{\bf{A}}_l} & {{\bf{B}}_l}  \\
   {{\bf{B}}_l^{\rm{H}}} & {\frac{c_l}{r}{\bf{I}}_r}  \\
\end{array}} \right]
\end{align} the homogenized optimization problem of (\ref{optimization_P_2}) is formulated as
\begin{align}
& \min \ \ {\rm{Tr}}({\bf{M}}(f_0)[{\bf{Y}};{\bf{Z}}][{\bf{Y}};{\bf{Z}}]^{\rm{H}}) \nonumber \\
& \ {\rm{s.t.}} \ \ \ {\rm{Tr}}({\bf{M}}(f_i)[{\bf{Y}};{\bf{Z}}][{\bf{Y}};{\bf{Z}}]^{\rm{H}}) \le \alpha_i, i \in {\mathcal I} \nonumber \\
& \ \ \ \ \ \ \ \ {\rm{Tr}}({\bf{M}}(f_j)[{\bf{Y}};{\bf{Z}}][{\bf{Y}};{\bf{Z}}]^{\rm{H}})=\alpha_j, j \in {\mathcal E} \nonumber \\
& \ \ \ \ \ \ \ \ {\bf{Z}}^{\rm{H}}{\bf{Z}}={\bf{I}}_r  \ \ \ \ {\bf{Y}}\in {\mathbb C}^{n\times r}.
\end{align}Notice that the optimal solution of (\ref{optimization_P_2}) ${\bf{X}}_{\rm{opt}}$ equals ${\bf{X}}_{\rm{opt}}={\bf{Y}}_{\rm{opt}}{\bf{Z}}_{\rm{opt}}^{\rm{H}}$. Defining ${\bf{U}}\triangleq [{\bf{Y}};{\bf{Z}}][{\bf{Y}};{\bf{Z}}]^{\rm{H}}$, after relaxing the rank constraint on ${\bf{U}}$, we have the following optimization problem
\begin{align}
\label{T-2-QMP}
& \min_{\bf{U}} \ \ {\rm{Tr}}({\bf{M}}(f_0){\bf{U}}) \nonumber \\
& \ {\rm{s.t.}} \ \ \ {\rm{Tr}}({\bf{M}}(f_i){\bf{U}}) \le \alpha_i, i \in {\mathcal I} \nonumber \\
& \ \ \ \ \ \ \ \ {\rm{Tr}}({\bf{M}}(f_j){\bf{U}})=\alpha_j, j \in {\mathcal E} \nonumber \\
& \ \ \ \ \ \ \ \ [{\bf{U}}]_{n+1:n+r,n+1:n+r}={\bf{I}}_r \ \ \ \ {\bf{U}}\succeq {\bf{0}}.
\end{align}
To recover ${\bf{X}}$ from ${\bf{U}}$, an algorithm based on rank reduction has been discussed in detail in \cite{Beck07}. When the number of the constraints are less than $2r$, this relaxation is tight \cite{Beck09}. Comparing (\ref{T-2-QMP}) with (\ref{SDR}), it can be observed that the SDP problem for T-2-QMP problems has a much lower dimension than that for T-1-QMP. It is because the T-2-QMP problems have a better structure to be exploited. In other words, it can be concluded that T-2-QMP problems have  much stronger convexity than T-1-QMP problems.

\noindent \textbf{Applications:} The solution of the T-2-QMP problem can be applied to AF MIMO relaying transceiver design at the source node with cognitive radio interference constraints.

\subsection{Discussions}
As mentioned at the beginning of this section, QMP is a special case of quadratically constrained quadratic programming (QCQP) discussed in \cite{ZQLuo2010}, it is important to compare the QMP-based algorithms with the QCQP-based algorithms given in \cite{ZQLuo2010}. Due to the fact that QMP is a special case of QCQP, QMP problems have better structures and enjoy better properties. For example, for the general T-2-QMP problems, they have stronger duality in semidefinite relaxation than the QCQP problems discussed in \cite{ZQLuo2010}. Particularly, for the case when there is only one constraint, using the QMP-based algorithm, the optimal solution can be computed by using a bisection search instead of solving  a SDP problem.  On the other hand, solving T-2-QMP problems, QMP-based algorithms have a much smaller dimension QCQP-based algorithms. Based on the complexity analysis in \cite{Vandenberghe96}, if the matrix variable ${\bf{X}}$  is an $M\times M$ matrix, for T-2-QMP problems using QMP-based algorithm in (\ref{T-2-QMP}) the complexity is $\mathcal{O}(M^{3.5}{\rm{ln}}(1/\epsilon))$ where $\epsilon$ is the precision. While using the QCQP-based algorithm in \cite{ZQLuo2010} the complexity is $\mathcal{O}(M^7{\rm{ln}}(1/\epsilon))$. It can be seen that the QMP-based algorithms have a great advantage in terms of computational complexity.

In addition, it is also very interesting to compare the QMP-based algorithms with the brute force iterative algorithms in which matrix variables are just taken as multi-dimensional vector variables and then brute force algorithms such as neural network algorithms are used to compute them. The main advantage of the QMP-based algorithms is that for QMP-based algorithms  some nature of the optimization problems is revealed and this is the reason why in certain cases even with a constraint, the solution has a semi-closed-form solution. For the general cases, the QMP-based algorithms can exploit the problem structure to improve the precision of the final solution and accelerate the convergence speed of the algorithm.

\section{Robust Transceiver Designs Based on QMP}
\label{Sect_5}

From the practical viewpoint, due to the limited length of training sequences and time varying nature of wireless channels, channel estimation errors are always inevitable. Channel errors will significantly decreases system performance. It is well-established that robust transceiver designs or beamforming designs can mitigate this negative effects \cite{Xing1012,XingICASSP2010}. A question naturally arises that whether the previously discussed QMP-based algorithms can be applied to so-called robust transceiver designs. This is exactly the focus of this section.

When channel errors are considered, the channel state information can be written as \cite{Xing1012}
\begin{align}
{\bf{H}}_{l}&={\bf{\bar H}}_{l}+\Delta{\bf{H}}_{l}
\end{align}where ${\bf{\bar H}}_{l}$ is the estimated ${\bf{H}}_{l}$ and $\Delta{\bf{H}}_{l}$ is the corresponding channel estimation error, respectively. The kronecker correlation model is widely used for channel estimation errors \cite{Xing1012,XingICASSP2010}
\begin{align}
\label{kronecker}
\Delta{\bf{H}}_{l}={\boldsymbol{\Sigma}}_l^{\frac{1}{2}}
{\bf{H}}_{W,l}{\boldsymbol{\Psi}}_l^{\frac{1}{2}}.
\end{align}where ${\boldsymbol{\Sigma}}_l$ and ${\boldsymbol{\Psi}}_l$ are the row and column correlation matrices, respectively. The inner matrix ${\bf{H}}_{W,l}$ is a random matrix with i.i.d Gaussian random elements with zero mean and unit variance.
Take the simplest point-to-point MIMO system as example to illustrate the impact of random matrix integrations.
For the point-to-point MIMO system, the data MSE at the destination equals to \cite{Palomar03}
\begin{align}
{\mathbb{E}}\{{\rm{Tr}}({\bf{G}}{\bf{H}}{\bf{F}}{\bf{F}}^{\rm{H}}
{\bf{H}}^{\rm{H}}{\bf{G}}^{\rm{H}})
-2{\mathcal{R}}\{{\rm{Tr}}({\bf{G}}{\bf{H}}{\bf{F}})\}+\sigma_n^2{\rm{Tr}}({\bf{G}}{\bf{G}}^{\rm{H}})
\}
\end{align}where the expectation operation at the outside is due to channel estimation errors. This equation is a QM function with respect to ${\bf{F}}$ or ${\bf{G}}$.
As a QM function consists of zero order term, first order term and second order term of the variables, in the following the matrix integrations over them are discussed separately. Zero-term is a constant and it is obvious that its integration with respect to any variable is itself.

Notice that the channel estimation errors are independent of the signal and the noise and their means are all zero. Based on these facts we directly have the following result for the first order term
\begin{align}
{\mathbb{E}}\{{\bf{H}}_{l}{\bf{X}}\}&={\bf{\bar H}}_{l}{\bf{X}}.
\end{align}The integration over the second order term is a little bit complicated. In order to make it clear, a preliminary result on complex matrix integration is given first.

\noindent \underline{\textbf{Complex matrix integration:}} For two $M \times N $ random complex matrices ${\bf{Q}}$ and ${\bf{W}}$, if they satisfy
\begin{align}
{\mathbb{E}}\{ {\rm{vec}}({\bf{Q}}){\rm{vec}}^{\rm{H}}({\bf{W}})
\}={\bf{A}}\otimes {\bf{B}},
\end{align} the following equality holds
\begin{align}
{\boldsymbol{\Sigma}}={\mathbb{E}}\{
{\bf{Q}}{\bf{R}}{\bf{W}}^{\rm{H}} \}={\bf{B}}{\rm{Tr}}({\bf{R}}{\bf{A}}^{\rm{T}})
\end{align}

\noindent \textbf{Proof:} See Appendix~A. $\blacksquare$

Based on the Kronecker product model (\ref{kronecker}) and the preliminary result we have the following equation
\begin{align}
&{\mathbb{E}}\{{\bf{H}}_{l}{\bf{X}}{\bf{X}}^{\rm{H}}{\bf{H}}_l^{\rm{H}}\}={\bf{\bar H}}_{l}{\bf{X}}{\bf{X}}^{\rm{H}}{\bf{\bar H}}_l^{\rm{H}}+{\rm{Tr}}({\bf{X}}{\bf{X}}^{\rm{H}}{\boldsymbol \Psi}_l){\boldsymbol \Sigma}_l.
\end{align}It is obvious that the expectation of a second-order term is also a second-order term. The main difference compared to the perfect case is that there is a residual part ${\rm{Tr}}({\bf{X}}{\bf{X}}^{\rm{H}}{\boldsymbol \Psi}_l){\boldsymbol \Sigma}_l$ caused by channel error. Based on the results on the expectation on the zero term, first order term and the second order term we have the following conclusion.

\noindent \underline{\textbf{Conclusion 3:}} For LMMSE transceiver designs, expectations of channel estimation errors keep the quadratic nature of the original QMP problems. Then it is not surprising that QMP technology can also be used in robust transceiver designs.

\noindent \textbf{Remark:} In the reference \cite{Gupta00}, only the matrix operations for real matrix variates are presented. Strictly speaking, it is not rigorous to directly use the results in that book \cite{Gupta00} or simply replace the symbol ${\rm{T}}$ by the symbol ${\rm{H}}$ in the involved matrix operations. Here for completeness we give a detailed proof about complex matrix integrations to make sure our results are rigorous.


\section{Numerical Results}
\label{Sect_6}
In this simulation part, in order to assess the effectiveness of the proposed solution, two different examples are shown. In the first example, there are two pairs of source and destination. Moreover, there are two relays facilitating the communications between the sources and their corresponding destinations. The direct links between the sources and destinations are neglected due to deep fading. In Example 1, the source nodes only transmit signals and the destination nodes only receive signals. In the second example, there are two sources to exchange information assisted by two relays. In order to improve the spectral efficiency, the famous physical layer network coding strategy named two-way relaying is adopted. Specifically, in the first time slot, two source terminals send their information to the relays and then the relays broadcast the filtered received signals to the two terminals. After that each terminal removes its own transmitted signal in the first time slot first and then recovers its desired signal.

In both the two examples, all nodes are equipped with multiple antennas. At each source node, two independent data streams, each with 10000 independent quadrature phase-shift keying (QPSK) symbols, are transmitted. Each point in the following figures is an average over 500 independent channel realizations. Furthermore, the famous Matlab toolbox CVX \cite{Grant07} is used in this paper to solve the standard convex optimization problems.

\noindent \underline{\textbf{Example 1:}}

In Example 1 for simplicity all nodes are equipped with $N_t$ antennas. In the first hop, the noise covariance matrices at the two relays are defined as ${\bf{R}}_{{\bf{n}}_{1,1}}$ and ${\bf{R}}_{{\bf{n}}_{1,2}}$, respectively. Without loss of generality, it is assumed that ${\bf{R}}_{{\bf{n}}_{1,1}}={\bf{R}}_{{\bf{n}}_{1,2}}
=\sigma_{n_1}^2{\bf{I}}_{N_t}$. Similarly, in the second hop, the noise covariance matrices at different destination are defined as ${\bf{R}}_{{\bf{n}}_{2,1}}
={\bf{R}}_{{\bf{n}}_{2,2}}=\sigma_{n_2}^2{\bf{I}}_{N_t}$. The signal-to-noise ratios (SNRs) for the source-relay links are defined to be ${\rm{E}}_{sr,k}=P_{s,k}/N_t\sigma_{n_1}^2$ , and are fixed to be ${\rm{E}}_{sr,k}=20{\rm{dB}}$.  The SNR for each relay-destination link is defined as ${\rm{E}}_{rd,k}=P_{r,k}/N_t\sigma_{n_2}^2$.

\begin{figure}[!ht]
\centering
\includegraphics[width=.7\textwidth]{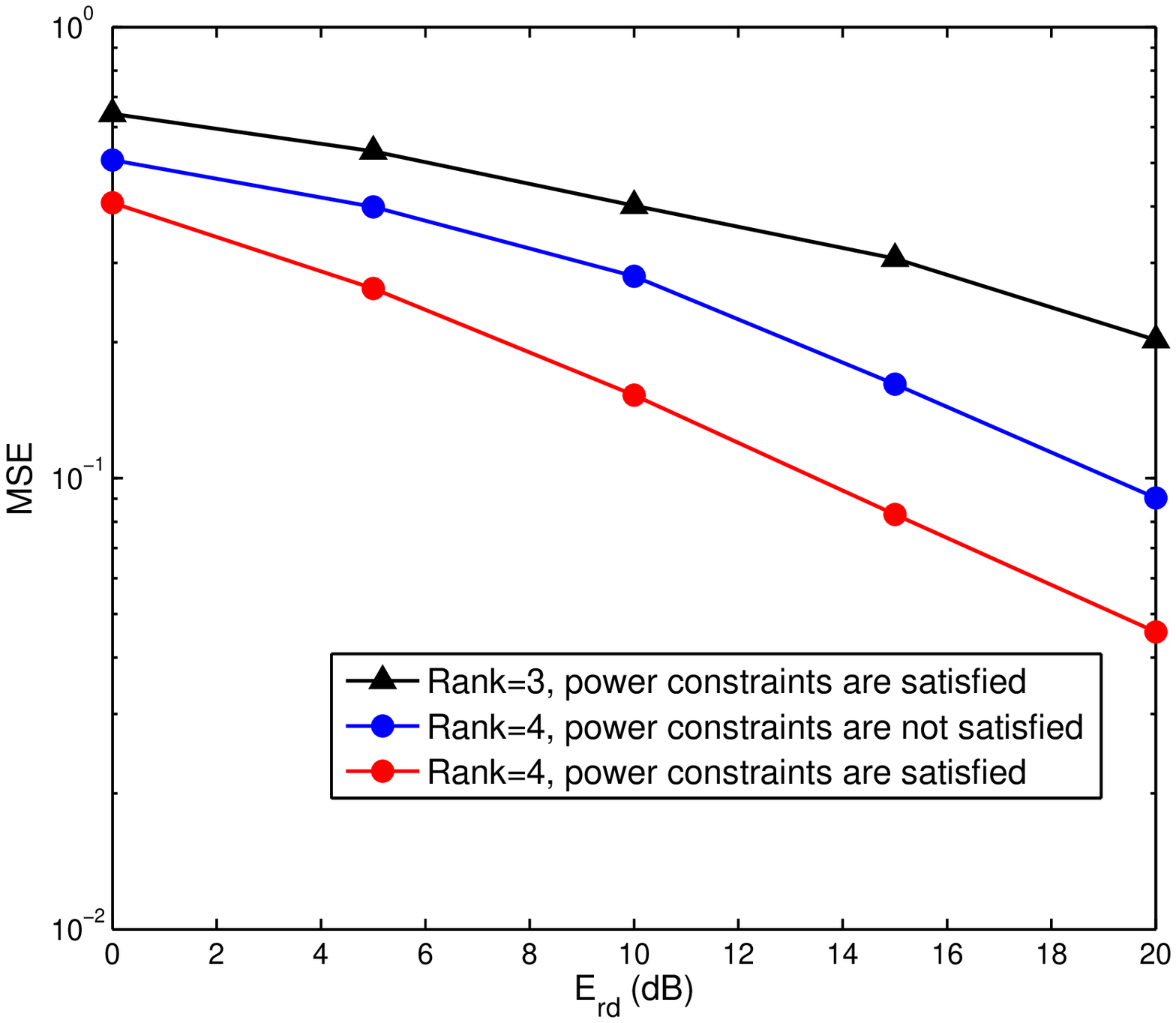}
\caption{Averaged MSE performance of the proposed algorithm with different initial values in Example 1.}\label{fig:2}
\end{figure}

For iterative algorithms, there is a well-known criterion for  the initial points selection. It states that the initial value should be close to the optimal solution. However, this criterion seems  meaningless as the optimal solution is usually unknown. Fig.~\ref{fig:2} shows the total data detection MSEs of the proposed algorithm with different initial precoder matrices at the source and relay when $N_t=4$. In our simulation settings, three kinds of initial values are selected to make a comparison, i.e., full rank identity matrix with the power constraints satisfied, full rank identity matrix without the constraints satisfied, diagonal matrices with rank of 3 and satisfying the power constraints. It can be observed that the initial values being full rank are much better than that with lower rank. The reason is full rank initial values can provide a larger available set for the following optimal value search than the lower rank initial values. Furthermore, for the full rank initial values, the one satisfying constraints is better than that without satisfying constraints. As for most of practical transceiver designs, the optimal solutions always occur on the boundary of the constraints. As a result the initial values satisfying constraints seem to be much closer than those without satisfying constraints and then they have better performance.

\begin{figure}[!ht]
\centering
\includegraphics[width=.7\textwidth]{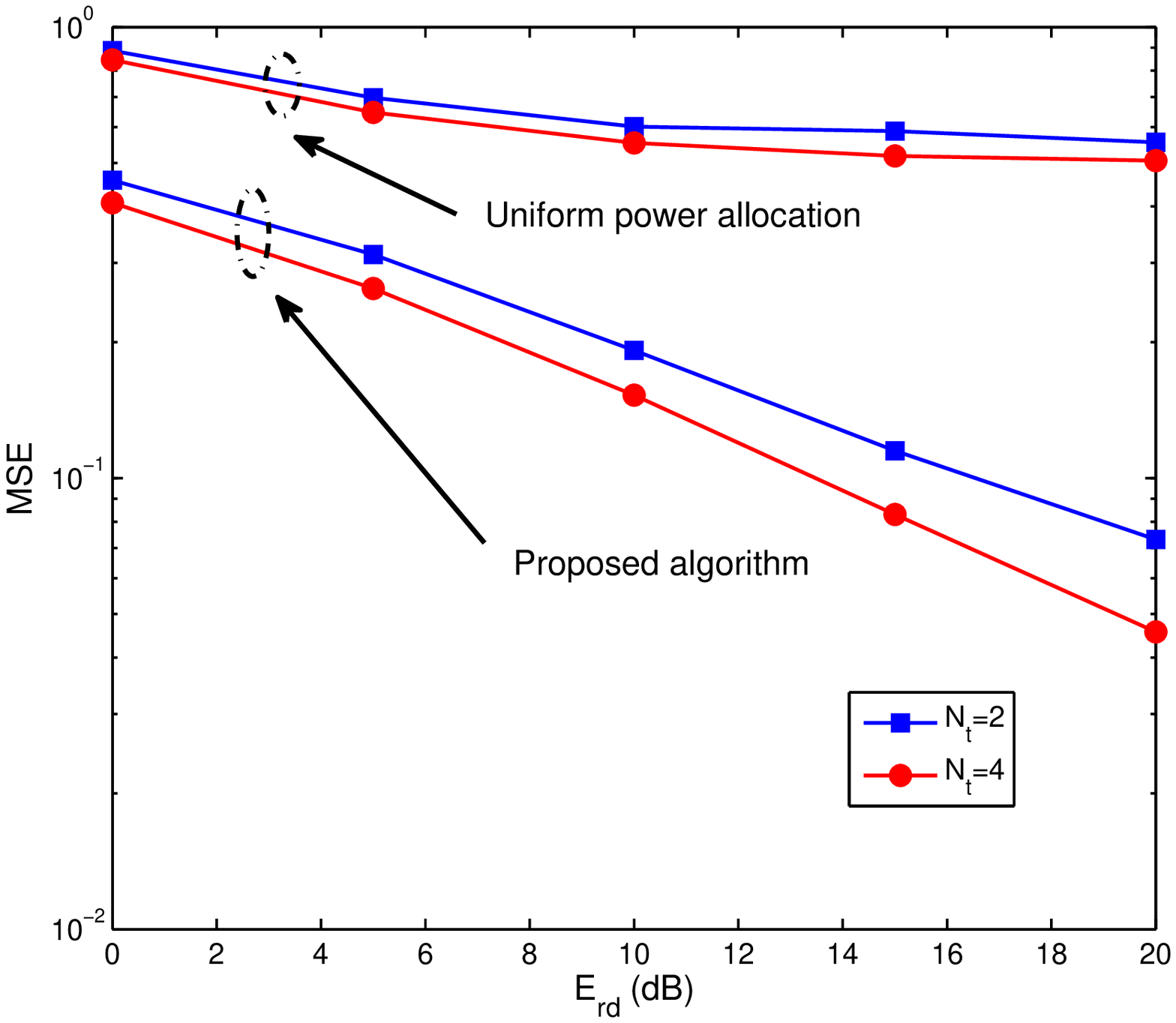}
\caption{Comparisons between the proposed algorithm and the uniform power allocation scheme in Example 1.}\label{fig:3}
\end{figure}

Fig.~\ref{fig:3} shows the performance advantage of the proposed algorithm over  the simplest uniform power allocation scheme in terms of averaged MSE in the two different cases $N_t=2$ and $N_t=4$. In uniform power allocation algorithm, the precoder matrices at the sources and relay are proportional to the identity matrices which are scaled by factors to make the equalities in the power constraints. In conclusion we can say that the proposed iterative algorithm can act as a better benchmark algorithm compared with the naive uniform power allocation scheme.

\noindent \underline{\textbf{Example 2:}}

In Example 2, the sources equipped with two antennas, i.e., $N_s=2$.  The two relays  are equipped with $N_r$ antennas. The noise covariance matrices at the relays are set as ${\bf{R}}_{{\bf{n}}_{r,1}}={\bf{R}}_{{\bf{n}}_{r,2}}
=\sigma_{n_r}^2{\bf{I}}_{N_r}$. Similarly at the sources, the noise covariance matrices are ${\bf{R}}_{{\bf{n}}_{s,1}}={\bf{R}}_{{\bf{n}}_{s,2}}
=\sigma_{n_s}^2{\bf{I}}_{N_s}$. Then in the first time slot SNRs for the source-relay links in the first slot are defined as ${\rm{E}}_{sr,k}=P_{s,k}/N_s\sigma_{n_r}^2$ and fixed to be 20dB. In the second time slot, the SNR for each relay-destination link is defined as ${\rm{E}}_{rs,k}=P_{r,k}/N_r\sigma_{n_s}^2$, and without loss of generality, it is assumed that ${\rm{E}}_{rs,1}={\rm{E}}_{rs,2}={\rm{E}}_{rs}$.

\begin{figure}[!ht]
\centering
\includegraphics[width=.7\textwidth]{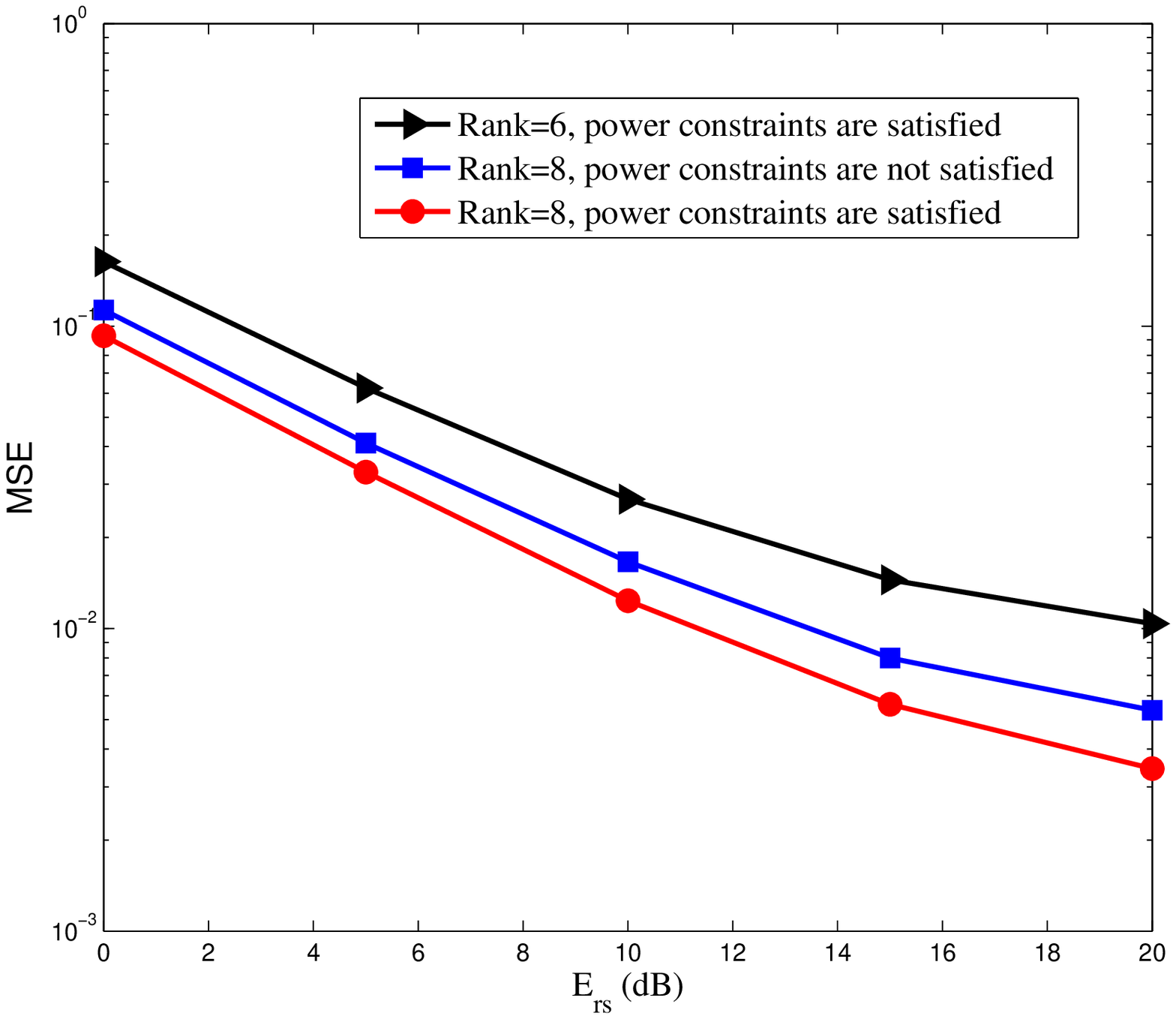}
\caption{Averaged MSE performance of the proposed algorithm with different initial values in Example 2.}\label{fig:4}
\end{figure}

The total MSEs of the proposed algorithm with different initial precoder matrices at the relays with $N_r=8$ are shown in Fig.~\ref{fig:4}. A similar result to Example 1 is achieved. In the two-way relaying network, the full rank initial value satisfying the constraints leads to the best performance and the ill-rank initial value with rank being 6 is the worst one.

\begin{figure}[!ht]
\centering
\includegraphics[width=.7\textwidth]{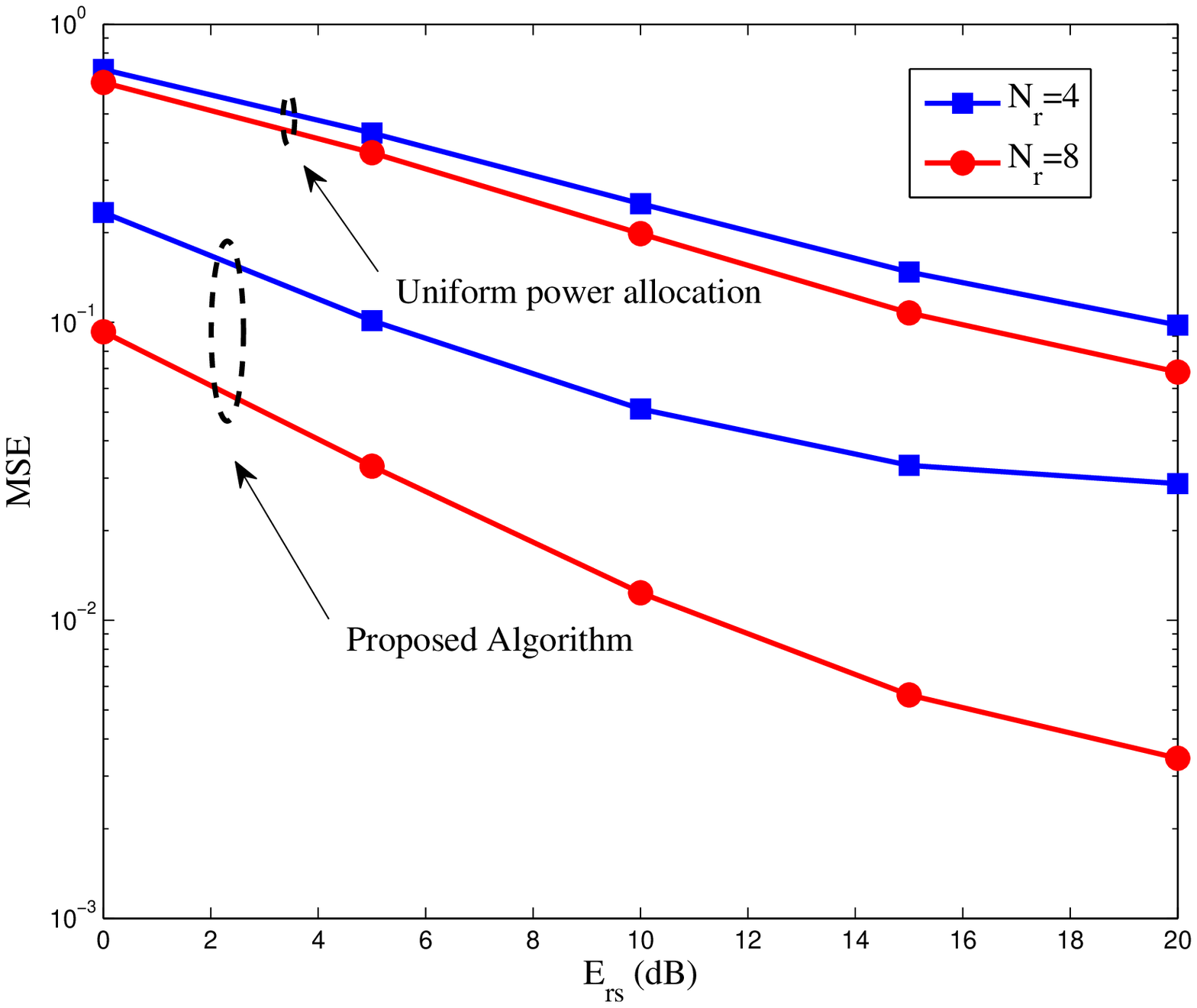}
\caption{Comparisons between the proposed algorithm and the uniform power allocation scheme in Example 2.}\label{fig:5}
\end{figure}

In Fig.~\ref{fig:5}, we compare the total MSEs of the uniform power allocation strategy and proposed algorithm in cases of $N_r=4$ and $N_r=8$. It is shown that for the two-way relaying network, the proposed iterative algorithm also performs much better than the uniform power allocation strategy. By the way as the number of antennas at the relay increases, the performance advantage of the proposed algorithm  becomes larger. Both Examples 1 and 2 have demonstrated the effectiveness of our proposed iterative algorithm and verified the correctness of our theoretical analysis.

\section{Conclusions}

In this paper, we discussed a unified iterative linear transceiver design with MSE as the performance criterion for different wireless systems. Different from the previous existing work, in our work the transceiver designs were understood from a unified optimization problem named as QMP problems for various wireless systems. The QMP-based designs can be applied to multi-cell coordinated beamforming designs, multi-user MIMO beamforming designs, cognitive radio MIMO beamforming designs, beamforming designs for cooperative networks and their robust designs with Gaussian random distributed channel estimation errors with row and column correlations. Along with transceiver designs, the elegant properties of QMP problems were also discussed in detail. In addition a framework on how to solve QMP problems was also given. The work presented in this paper will act as a baseline algorithm for the future wireless transceiver designs.

\section*{Acknowledgment}

This work is supported in part by the National Natural
Science Foundations of China (NSFC) under Grant No. 61101130.

\appendices


\section{Complex Matrix Integration}
\label{Lemma}

For the expectation of the following product
\begin{align}
{\boldsymbol{\Sigma}}={\mathbb{E}}\{
{\bf{Q}}{\bf{R}}{\bf{W}}^{\rm{H}} \}
\end{align} where ${\bf{Q}}$ and ${\bf{W}}$
are two $M \times N $ random matrices with compatible dimension to
${\bf{R}}$, the $(i,j)^{\rm{th}}$ element of ${\boldsymbol{\Sigma}}$
is
\begin{align}
\label{sigma}
[{\boldsymbol{\Sigma}}]_{i,j}=&{\mathbb{E}}\{[{\bf{Q}}]_{i,:}{\bf{R}}[{\bf{W}}]_{j,:}^{\rm{H}}\}\nonumber \\
=&\sum_{t}\sum_{k}{\mathbb{E}}\{[{\bf{Q}}]_{i,t}[{\bf{R}}]_{t,k}[{\bf{W}}]_{j,k}^*\}.
\end{align}
If the two random matrices ${\bf{Q}}$ and ${\bf{W}}$ satisfy
\begin{align}
{\mathbb{E}}\{ {\rm{vec}}({\bf{Q}}){\rm{vec}}^{\rm{H}}({\bf{W}})
\}={\bf{A}}\otimes {\bf{B}},
\end{align} where ${\bf{A}}$ is a $N \times N$ matrix while
${\bf{B}}$ is a $M \times M$ matrix, then we have the
equality $
{\mathbb{E}}\{[{\bf{Q}}]_{i_1,j_1}[{\bf{W}}]_{i_2,j_2}^{*}\}=[{\bf{B}}]_{i_1,i_2}[{\bf{A}}]_{j_1,j_2}$.
As $[{\bf{Q}}]_{i,t}$ and $[{\bf{W}}]_{j,k}$ are scalars,
(\ref{sigma}) can be further written as
\begin{align}
\label{92}
[{\boldsymbol{\Sigma}}]_{i,j}=&\sum_{t}\sum_{k}([{\bf{R}}]_{t,k}{\mathbb{E}}\{[{\bf{Q}}]_{i,t}[{\bf{W}}]_{j,k}^*\})\nonumber \\
=&\sum_{t}\sum_{k}[{\bf{R}}]_{t,k}[{\bf{A}}]_{t,k}[{\bf{B}}]_{i,j}.
\end{align}Finally,
writing (\ref{92}) back to matrix form, we have \cite{Kay93}
\begin{align}
\label{Expectation}
{\boldsymbol{\Sigma}}&={\bf{B}}{\rm{Tr}}({\bf{R}}{\bf{A}}^{\rm{T}}).
\end{align} Notice that this conclusion is independent of the matrix
variate distributions of  ${\bf{Q}}$ and ${\bf{W}}$, but only
determined by their second order moments.


\end{document}